\documentclass{aastex631}

\usepackage{amsmath, amssymb, amsfonts}
\usepackage{float}
\usepackage{multirow}
\usepackage{subfigure}
\usepackage{graphicx}
\graphicspath{{./}{figures/}}

\begin{document}

\title{Search for Distant Hypervelocity Star Candidates Using RR Lyrae Stars}
\author{Haozhu Fu}
\affiliation{Department of Astronomy, School of Physics, Peking University, Beijing 100871, China}
\affiliation{Kavli Institute for Astronomy and Astrophysics, Peking University, Beijing 100871, China}
\author{Yang Huang}
\affiliation{School of Astronomy and Space Science, University of Chinese Academy of Science, Beijing 100049, China}
\affiliation{Key Lab of Optical Astronomy, National Astronomical Observatory, Chinese Academy of Sciences (CAS) A20 Datun Road, Chaoyang, Beijing 100101, China
}
\author{Huawei Zhang}
\affiliation{Department of Astronomy, School of Physics, Peking University, Beijing 100871, China}
\affiliation{Kavli Institute for Astronomy and Astrophysics, Peking University, Beijing 100871, China}

\correspondingauthor{Yang Huang and Huawei Zhang}
\email{huangyang@ucas.ac.cn and zhanghw@pku.edu.cn}

\begin{abstract}
    Hypervelocity stars (HVSs) are stars with velocities exceeding their local escape velocities. Searching for HVSs and studying their origins can be an important way to study the properties of the Milky Way. 
    In this paper, we utilize precise distances for RR Lyrae stars (RRLs) derived from the period-absolute magnitude-metallicity (PMZ) relation, along with proper motions from Gaia DR3, to conduct a large-volume search for HVSs. Our sample consists of a catalog with 8,172 RRLs with metallicity, distance and radial velocities estimated from SDSS and LAMOST spectroscopic data, and an extended catalog of 135,873 RRLs with metallicity and distance estimated from Gaia photometry.
    After careful quality cuts, 165 hypervelocity RRL candidates were found. We performed further checks on their light curves, and selected the most reliable 87 hypervelocity RRLs. All of them exceed the Milky Way's escape velocity in the tangential component. Among them, 7 stars have tangential velocity over 800 $\rm km\,s^{-1}$. We identified two spatially distinct distributions of hypervelocity RRLs: one concentrated toward the Galactic Center and another localized around the Magellanic Clouds, suggesting that their origins are likely associated with these regions through the Hills or other mechanisms. Furthermore, we detected a significant number of RRLs associated with dwarf galaxies that exceed the Milky Way’s escape velocity, likely ejected from their host systems. Future Gaia releases and spectroscopic follow-up observations will provide further insight into their ejection origin. 
\end{abstract}

\keywords{Hypervelocity stars (776) --- RR Lyrae variable stars (1410) --- Milky Way dynamics (1051) --- Galaxy structure (622) --- Distance indicators (394) }


\section{Introduction} \label{sec:intro}
Hypervelocity stars (HVSs) are stars whose velocities exceed the local escape velocity, making them gravitationally unbound from the Milky Way. Researches on high-velocity stars began in the 1920s, focusing on the proper motion of stars \citep{vanMaanen_1917}. 
Subsequently, radial velocity measurements from spectroscopic observations revealed several OB-type stars in the Galactic halo with velocities up to 260 $\rm km\,s^{-1}$ \citep{1974ApJS...28..157G}. Owing to their youth and unusually high speeds---atypical of halo stars---they were inferred to have been ejected from their birthplaces into the halo, and were thus termed ``runaway stars.''

In 1988, Hills first proposed a mechanism capable of accelerating stars to velocities of up to 1000 km s$^{-1}$ or even faster \citep{1988Natur.331..687H}. When a binary system approaches the supermassive black hole (SMBH) at the Galactic Center (GC), the SMBH’s tidal force can disrupt the system. One of the binary companions can be captured by the SMBH, while the other one can be ejected at speeds reaching 1000 km s$^{-1}$, exceeding the escape velocity of the GC. In this scenario, the ejected stars are referred to as hypervelocity stars (HVSs).
In this paper, we generally define HVSs as stars that are unbound from the Milky Way. Later, \cite{Brown_2005} discovered the first HVS with a velocity of up to 700 km s$^{-1}$ and a Galactocentric distance between 40 and 70 kpc, exceeding the local escape velocity. This finding provided the first observational verification of the Hills mechanism.
Their following surveys for HVSs in the Galactic halo are also consistent with GC origin \citep{2014ApJ...787...89B}, as well as these early-type HVSs discovered by LAMOST \citep{Zheng_2014,2017ApJ...847L...9H,Li_2018,2025arXiv250414836S}. Furthermore, a HVS with velocity of approximately 1700 $\rm km\,s^{-1}$ was discovered to be confidently traced back from the GC \citep{2020MNRAS.491.2465K}.

Alternative ejection mechanisms, other than the Hills mechanism, can also explain the origin of HVSs. If they originate from the GC, they can also be produced by: (1) scattering by possible binary black hole at the GC \citep{2003ApJ...599.1129Y}, (2) interactions between the SMBH and accreted globular clusters \citep{2015MNRAS.454.2677C}. 
In addition to originating from associations with the SMBH at the GC, HVSs can be produced by (1) type Ia supernovae in close binary systems \citep{1961BAN....15..265B}, especially for white dwarf (WD)\text{–}WD binaries \citep{2018ApJ...865...15S} or binaries consisting of a WD and a non-degenerate star \citep{2009A&A...508L..27W}, which can eject HVS with velocity similar to the Hills mechanism, (2) scattering in dense stellar systems \citep{2009MNRAS.396..570G}, (3) tidal disruption of Milky Way's satellite galaxies \citep{2009ApJ...691L..63A}. Later searches for HVSs found supporting evidences for these scenarios. A hot subdwarf \citep{2015Sci...347.1126G} and some white dwarfs \citep{2018ApJ...865...15S,2023OJAp....6E..28E} were found to be HVSs originating from type Ia supernovae explosions. Furthermore, some HVSs were found to originate from various substructures of the Milky Way, such as dwarf galaxies \citep{Huang_2021,2022arXiv220704406L}. Most recent, a high-velocity star was found to originate from association with the intermediate-mass black hole (IMBH) at the center of a globular cluster \citep{2025NSRev..12..347H}.

Searching for HVSs and studying their origins provide valuable insights into the properties of the Milky Way. In particular, HVSs originating from the GC can help constrain the position and velocity of the Sun within the Milky Way \citep{2018ApJ...869...33H,2020MNRAS.491.2465K}, as well as refine our understanding of the distribution of stars and the parameter space of possible black hole companions around the GC \citep{2022MNRAS.512.2350E}. Additionally, HVSs serve as useful tracers for constraining the distribution and potential of the Milky Way’s dark matter halo \citep{2005ApJ...634..344G}.

In the search for HVSs, determining the distance and velocity of the stars is crucial to assess whether they can exceed their local escape velocity. However, the challenge lies in the fact that the extreme speeds of HVSs can result in them traveling significant distances from their origins, while the volume we search for HVSs is strongly constrained by accurate distance measurements.

Thus, one method for identifying HVSs is by selecting early-type stars with extreme radial velocities in spectroscopic surveys towards the Galactic halo, as is performed by \cite{2014ApJ...787...89B}. These stars are bright and thus easy to observe at large volume. Additionally, early-type stars are not typically found in the Galactic halo, their presence suggests they may have been ejected from the Galactic disk or the GC, likely with high velocities. However, early-type massive stars represent only a small fraction of the stellar population, whereas the majority of stars are late-type, low-mass stars, according to the initial mass function. These stars are too faint to be observed at large distances in spectroscopic surveys, which makes detecting late-type HVSs based solely on extreme radial velocities a challenging task.

Another method to search for HVSs is selecting stars with high tangential velocity based on astrometric measurements. The Gaia satellite, launched in 2013, provides a huge number of accurate astrometric measurements, especially the recently released Gaia DR3 \citep{gaiacollaboration2022gaia}. Some HVSs have been discovered based on parallax and proper motion provided by Gaia DR3 \citep{2023ApJ...944L..39L,2022MNRAS.515..767M}. However, due to the limitation in parallax measurement, accurate distance data for faint, distant stars remains challenging. For instance, the median parallax uncertainties of Gaia DR3 are $0.02\,\text{–}\,0.03$ mas for $G<15$; 0.07 mas at $G=17$ and 0.5 mas at $G=20$, corresponding to distance limits of $\sim40$, $14$ and $2$ kpc respectively. 

To address this issue, we utilize standard candles such as RR Lyrae variable stars to measure distances up to 100 kpc. Given the lack of precise radial velocities, we focus on stars with large tangential velocities. By incorporating accurate proper motion measurements from Gaia DR3, we can significantly expand the search volume for HVSs.

RR Lyrae variables (RRLs) are old, metal-poor, helium-burning giants located in the instability strip of the Hertzsprung-Russell (H-R) diagram. Their typical pulsation periods range from 0.2 to 1 day, with V-band amplitudes reaching up to 1.5 magnitudes. From their light curves, both metallicity and absolute magnitude can be derived \citep{Drake_2014,2023ApJ...944...88L}. As a result, their distances can be precisely measured for large samples, even when only time-domain photometry is available.

Previous work has used RRLs to search for HVSs \citep{Prudil_2022}. Their sample has 6,187 RRLs compiled from Gaia DR2 \citep{2019A&A...622A..60C} and the Catalina Sky Survey \citep{2014yCat..22130009D} with 3D velocity data. 
In this paper, we utilize a sample of approximately 140,000 RRLs from \cite{Wang_2022} and \cite{2023ApJ...944...88L}, combined with proper motion measurements from Gaia DR3, to conduct a large-scale search for hypervelocity RR Lyrae candidates (HV-RRLs), with a particular focus on those exhibiting high tangential velocities.
The selection process of HV-RRL candidates will be shown in Section \ref{sec:metho}; we will show the results and find their possible origins by integrating their backward orbits in Section \ref{sec:resul}; we will discuss the RRLs as member stars of the Milky Way's substructures in Section \ref{sec:discu}. Finally, a summary is presented in Section \ref{sec:summa}.

\section{Sample and Method} \label{sec:metho}
\subsection{Sample selection}
In this study, we adopt two samples of RRL stars, which are introduced below.

\citet[hereafter W22]{Wang_2022} enlarged the RRL catalog derived by \cite{2020ApJS..247...68L}, which is based on General catalog of Variable Stars (GCVS, \citealt{10.1086/649432}), Catalina sky survey (CSS, \citealt{2014yCat..22130009D}) and other surveys. 8,172 RRLs are found to have low-to-medium resolution spectra collected by LAMOST Galactic spectroscopic surveys \citep{2012RAA....12..735D,2014IAUS..298..310L} and SDSS/SEGUE \citep{2009AJ....137.4377Y}. The metallicity and radial velocity of these RRLs are then estimated using the techniques developed in \cite{2020ApJS..247...68L}.

\citet[hereafter L23]{2023ApJ...944...88L} calibrated the relation between RRL's Fourier parameters of light curves and metallicity, based on spectroscopic metallicity estimates from \cite{2020ApJS..247...68L}. Then, L23 applied this relation to a much larger RRL catalog based on Gaia DR3 time-series photometry derived by \cite{clementini2022gaia}. As a result, 135,873 RRLs with precise photometric metallicity determination are selected. 

Finally, after removing duplicates, their catalogs provide key parameters for a total of 138,857 RRLs, including right ascension (RA), declination (Dec), metallicity, distance, absolute magnitude, and period. The typical uncertainty in metallicity is approximately 0.3 dex, while distance uncertainty is approximately 10\%. 

We then cross-match our sample with Gaia DR3 to retrieve proper motion data $(\mu^*_{\alpha}, \mu_{\delta})$. The typical uncertainties are $0.02\text{--}0.04$, $0.07\text{--}0.1$, and $0.5\text{--}1.5$ mas yr$^{-1}$ for $G < 15$, $G \approx 17$, and $G \sim 20\text{--}21$, respectively. Additionally, 8,561 RRLs in the catalogs have radial velocity measurements from LAMOST, SDSS spectroscopic data, or Gaia DR3. The typical radial velocity uncertainties range from $5$ to $21$ km s$^{-1}$. Additionally, to ensure data quality, we exclude radial velocity measurements of 19 RRLs with uncertainties greater than 30 $\rm km\,s^{-1}$.

\subsection{Data quality control}\label{2.2}
First, proper motion errors can introduce significant velocity uncertainties due to the large distances involved. Therefore, stringent astrometric cuts are necessary to exclude unreliable proper motion measurements. Following previous studies \citep{Prudil_2022,2022MNRAS.515..767M}, we apply the following selection criteria based on Gaia DR3 astrometric parameters \citep{2021A&A...649A...2L}:

\begin{enumerate}
    \item[(1)]
    $\texttt{RUWE}<1.4$: this parameter reflects the quality of the single-star model fit to the astrometric observations;
    \item[(2)]
    $\texttt{ipd\_gof\_harmonic\_amplitude}<0.4$: this indicates the amplitude of changes in the goodness-of-fit in image parameter determination;
    \item[(3)]
    $\texttt{ipd\_frac\_multi\_peak}<2$: this indicates the amount of image asymmetry, to exclude possible binarity;
    \item[(4)]
    $\texttt{astrometric\_excess\_noise\_sig}\le2.0$: this indicates whether the object is astrometrically well-behaved;
    \item[(5)]
    $\texttt{visibility\_periods\_used}\ge13$: this ensures the selected variables are well-observed sources.
\end{enumerate}

\begin{figure*}
	\centering
	\includegraphics[width=0.9\textwidth]{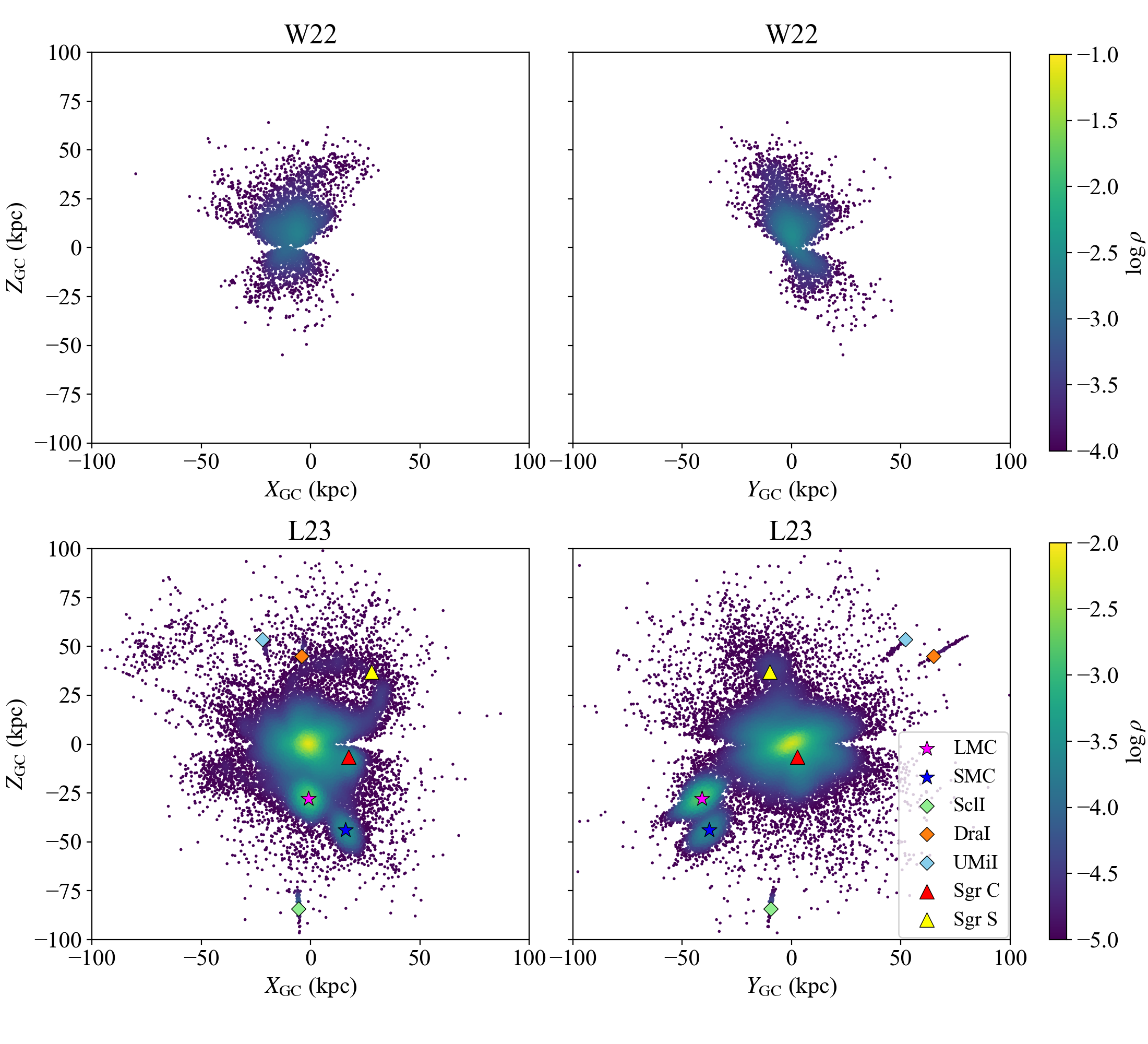}
	\caption{\label{fig5}Spatial distribution of RRL samples from W22 (upper panels) and L23 (bottom pannels) in the Galactic Cartesian coordinates. Left panels show $X_{\rm GC}\text{--}Z_{\rm GC}$ distribution; right panels show $Y_{\rm GC}\text{--}Z_{\rm GC}$ distribution. Color bar in the right side stands for number density in logarithmic scale. Magenta and blue stars denote the position of the LMC and SMC; Green, orange and skyblue diamonds denote the position of the Sculptor, Draco, and Ursa Minor dwarf galaxies (SclI, DraI, and UMiI); Yellow and Red triangles denote the Sgr core (C) and stream (S), respectively. The position of these dwarf galaxies are taken from \citet{2018A&A...619A.103F}.}
	\vspace{10pt}
\end{figure*}

After applying these criteria, 110,823 RRLs remain. Fig. \ref{fig5} presents the spatial distribution of the selected sample in Galactic Cartesian coordinates. Several substructures can be identified, as are marked in Fig. \ref{fig5}, including the Large Magellanic Cloud (LMC), the Small Magellanic Cloud (SMC), the stream of the Sagittraus spheroid dwarf galaxy (Sgr), as well as three other dwarf galaxies that appear elongated due to distance uncertainties. The position of these dwarf galaxies are taken from \citet{2018A&A...619A.103F}.

\cite{2024A&A...685A.162S} conducted a stringent examination on astrometric quality and provided a comprehensive review on HVS candidates from Gaia DR2 and DR3, with a particular focus on sources exhibiting unusually large proper motions.
To further ensure the astrometric quality of our sample, we applied the methodology outlined by \cite{2024A&A...685A.162S} to examine the astrometric flags and parameters provided by Gaia DR3 for each HV-RRL candidate. 
However, most stars in our sample with high tangential velocities exhibit relatively modest proper motions, with the large inferred velocities primarily driven by their substantial distances. As such, the criteria adopted by \citet{2024A&A...685A.162S} may not be fully applicable to our case.
To facilitate further use and verification, our final online catalog includes a flag indicating whether each candidate satisfies the selection criteria defined by \citet{2024A&A...685A.162S}.

\subsection{HVS selection}\label{2.3}
To select hypervelocity RRL candidates, we calculate each RRL's Galactocentric distance ($R_{\rm GC}$) and total velocity in the Galactic rest frame ($v_{\rm GC}$).
The Sun's coordinates in the Galactic rest frame are adopted as $R_{\odot} = 8.2$ kpc \citep{2016ARA&A..54..529B} and $Z_{\odot} = 20.8$ pc \citep{Bennett_2018}. The Galactocentric distance, $R_{\rm GC}$, is then calculated based on the RA, Dec, and distance $D$ provided by \cite{Wang_2022} and \cite{2023ApJ...944...88L}.
The Sun's peculiar motion relative to the Galactic rest frame is taken as $[U_{\odot}, V_{\odot}, W_{\odot}] = [11.1, 12.24, 7.25]$ km s$^{-1}$ \citep{2010MNRAS.403.1829S}, with a circular velocity at the solar radius of $v_{\rm c, R_0} = 234$ km s$^{-1}$ \citep{2023ApJ...946...73Z}. Additionally, we tested alternative values for the solar peculiar motion (e.g., \citealt{2015IAUGA..2251816H,2021MNRAS.504..199W}) and found that our results remain robust.

Next, we derive the total Galactic rest-frame velocity ($v_{\rm GC}$) for RRLs with available radial velocity ($v_r$) measurements. For RRLs lacking radial velocity data, we compute only their tangential velocity in the Galactic rest frame ($v_{{\rm GC},t}$). If $v_{{\rm GC},t}$ exceeds the local escape velocity, then the total velocity $v_{\rm GC}$ must be even higher, allowing these stars to be classified as HVS candidates.
The tangential velocity $v_{{\rm GC},t}$ is obtained by projecting both the star’s velocity and the Sun’s velocity perpendicular to the radial direction:
\begin{align}
    v_{{\rm GC},t}&=\left|\vec{v}_{{\rm obs},t}+\Big[\vec{v}_\odot-(\vec{v}_\odot\cdot\hat{l})\hat{l}\Big]\right|,
\end{align}
where $\vec{v}_{{\rm obs},t}$ is the observed tangential velocity, which equals to proper motion multiplying distance; $\vec{v}_\odot=(U_\odot,V_\odot+v_{c, R_0},W_\odot)$ is the velocity of the Sun in the Galactic rest frame and $\hat{l}$ indicates the unit vector along the line-of-sight. 

We conducted 2000 Monte Carlo simulations, neglecting the uncertainties in RA and Dec, while assuming that the remaining parameters ($\mu^*_{\alpha},\,\mu_{\delta},\,v_r,\,D$) follow a multi-dimensional Gaussian distribution, with the corresponding covariance matrix provided by the original catalogs and Gaia DR3. It is important to note that the distances are derived from the period–metallicity relation of RRLs rather than from parallax, and are therefore uncorrelated with proper motions. The final values of $R_{\rm GC}$ and $v_{\rm GC}$, along with their uncertainties ($\sigma_{R_{\rm GC}}$, $\sigma_{v_{\rm GC}}$), are determined from the 50th, 16th, and 84th percentiles of the 2000 Monte Carlo realizations.

To calculate the escape velocity ($v_{\rm esc}$) of the Milky Way, we utilized \texttt{MWPotential2014} in python package \texttt{galpy} \citep{2015ApJS..216...29B}. It's a potential distribution of the Milky Way consisting of a Miyamoto \& Nagai disk potential \citep{1975PASJ...27..533M}, a power-law bulge potential with an exponential cut-off and Navarro-Frenk-White spherical dark matter halo \citep{1997ApJ...490..493N}. In this work, escape velocity is calculated by $v_{\rm esc}=\sqrt{2(\Phi(\infty)-\Phi(r))}$, where $\Phi$ is the potential of the Milky Way. We then computed the probability ($P_{\rm unb}$) that each RRL has a total Galactic rest-frame velocity ($v_{\rm GC}$) exceeding the Milky Way's escape velocity ($v_{\rm esc}$). This was achieved by determining the fraction of samples where $v_{\rm GC} > v_{\rm esc}$ in 2000 times of Monte Carlo simulations. As a result, 8,090 RRLs with $P_{\rm unb}>0.5$ are found.

To ensure reliable velocity measurements, we exclude RRLs with large relative velocity uncertainties ($\sigma_{v_{\rm GC}} / v_{\rm GC} > 0.3$). A total of 6,838 RRLs meet this criterion.
These RRLs are classified into two groups based on their potential association with Milky Way substructures, including the LMC, SMC, Sgr stream/core, other well-known dwarf galaxies, and star clusters. Guided by \cite{2023arXiv230709572C}, we identify stars associated with these substructures using the criteria detailed below:

\begin{enumerate}
    \item[(1)]
    Located within the tidal radii in angular separation and within 5 kpc in line-of-sight distance from 145 Galactic globular clusters \citep{2010arXiv1012.3224H,2019yCat..74842832V,2021MNRAS.505.5978V,2021yCat..75055978V}.
    \item[(2)]
    Within 10 half-light radii in angular separation and within 20 kpc in line-of-sight distance from the centers of 39 dwarf galaxies
    \citep{2018A&A...619A.103F}.
    \item[(3)]
    The member stars of the LMC and SMC, which are defined by having close angular distances ($16^\circ$ and $12^\circ$) and small relative proper motions ($<\pm 15\,\rm mas\,yr^{-1}$) comparing to the LMC and SMC. Meanwhile, the RRLs in the LMC and SMC should have expected apparent magnitude of $G>18.5$ \citep{10.1093/mnras/stw3357}. The 6D data of the LMC and SMC is taken from \cite{2020ApJ...893..121P}.
    \item[(4)]
    Belong to the sample of RRL candidates from the Sgr Stream and Sgr Core, as identified by \cite{2020A&A...638A.104R}.
\end{enumerate}
\begin{deluxetable}{cccc}
\tabletypesize{\footnotesize}
\tablenum{1}
\tablecaption{\label{tab1}Number of RRLs after data quality control and HVS selection}
\tablehead{
selection   & W22   &  L23   & total}
\startdata
original                                   & 8172     &  135873 & 144045 \\
remove duplicates                           & 7652     &  131205 & 138857 \\
quality control                        & 5138     &  105685 & 110823 \\
$P_{\rm unb}>0.5$                       & 23       &  8067   & 8090 \\
$\sigma_{v_{\rm GC}}/v_{\rm GC}\le 0.3$ & 19       &  6819   & 6838\\
not in substructure                     & 19       &  146    & 165 \\
\enddata
\end{deluxetable}

As a result, we identified 6,673 unbound RRLs associated with various substructures of the Milky Way, constituting 97.6\% of the unbound RRLs after the HVS selection. Similar to the main Galaxy, these substructures can also produce HVSs via various mechanisms, including supernova explosions, multi-body scattering and Hills mechanism. \cite{2017MNRAS.469.2151B} conducted comprehensive simulations for the LMC and predicted that the runaway star production rate from the LMC could be as high as $3 \times 10^{-6}\,\rm yr^{-1}$. Recent simulations have also explored the ejection of HVSs from the LMC via the Hills mechanism. \cite{2021MNRAS.507.4997E} predicted that in future surveys, the number of observable HVSs ejected from the SMBH of the LMC will significantly exceed those originating from the GC, assuming a constant HVS ejection rate of $10^{-4}\,\rm yr^{-1}$. Furthermore, \cite{2025arXiv250200102H} proposed that the Leo overdensity of HVSs observed by \cite{2014ApJ...787...89B} can be attributed to the Hills mechanism in the LMC, with an ejection rate of $2\times10^{-6}\,\rm yr^{-1}$.

However, the selected unbound RRLs in these substructures may exceed the escape velocity simply due to additional velocity acquired from their host substructures, especially those already moving at high speeds in the Galactic rest frame. For instance, the LMC has an orbital velocity of approximately $330\,\rm km\,s^{-1}$, while the escape velocity at its position is around $360\,\rm km\,s^{-1}$.  More importantly, some of these unbound RRLs are located in dense star fields, where Gaia’s astrometric precision necessitates further examination \citep{2024gfpr.reptE....V}.

In this study, we first focus on the remaining unbound field stars that are not associated with any known substructures of the Milky Way, and subsequently discuss the unbound members associated with substructures in Section~\ref{dis:sub}. As a result, we selected a final sample of 165 hypervelocity RR Lyrae (HV-RRL) candidates. The data cuts and selection criteria are summarized in Table \ref{tab1}. 

\subsection{Light curve examination}\label{2.4}

Although we have applied strict astrometric cuts to ensure reliable proper motion data from Gaia DR3, the reliability of distance measurements in RRL catalogs remains uncertain. The accuracy of RRL distance measurements is influenced by the quality of their light curves and the precision of Fourier parameter determination. Additionally, different types of RRLs have distinct relationships for deriving metallicity and distance. In the catalogs of W22 and L23, RRLs are classified into RRab and RRc subtypes. RRab stars exhibit asymmetric light curves with a faster rise and slower decline, whereas RRc stars have more symmetric light curves resembling sine waves. Incorrect identification or classification of RRLs and their subtypes could lead to inaccurate distance measurements. Therefore, it is crucial to carefully examine the light curves of RRLs in the catalogs of W22 and L23.

\begin{figure*}
	\centering
	\subfigure{\includegraphics[width=0.32\linewidth]{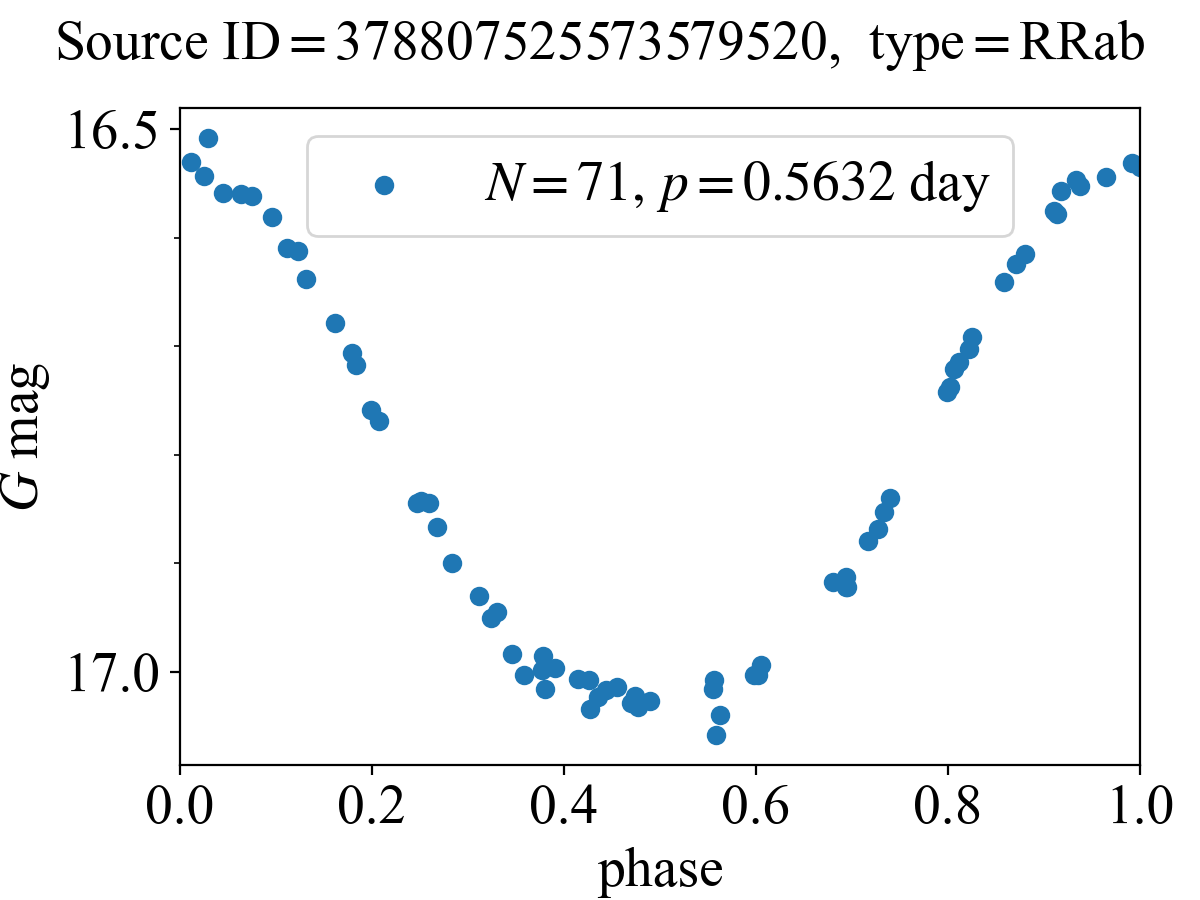}}
	\subfigure{\includegraphics[width=0.32\linewidth]{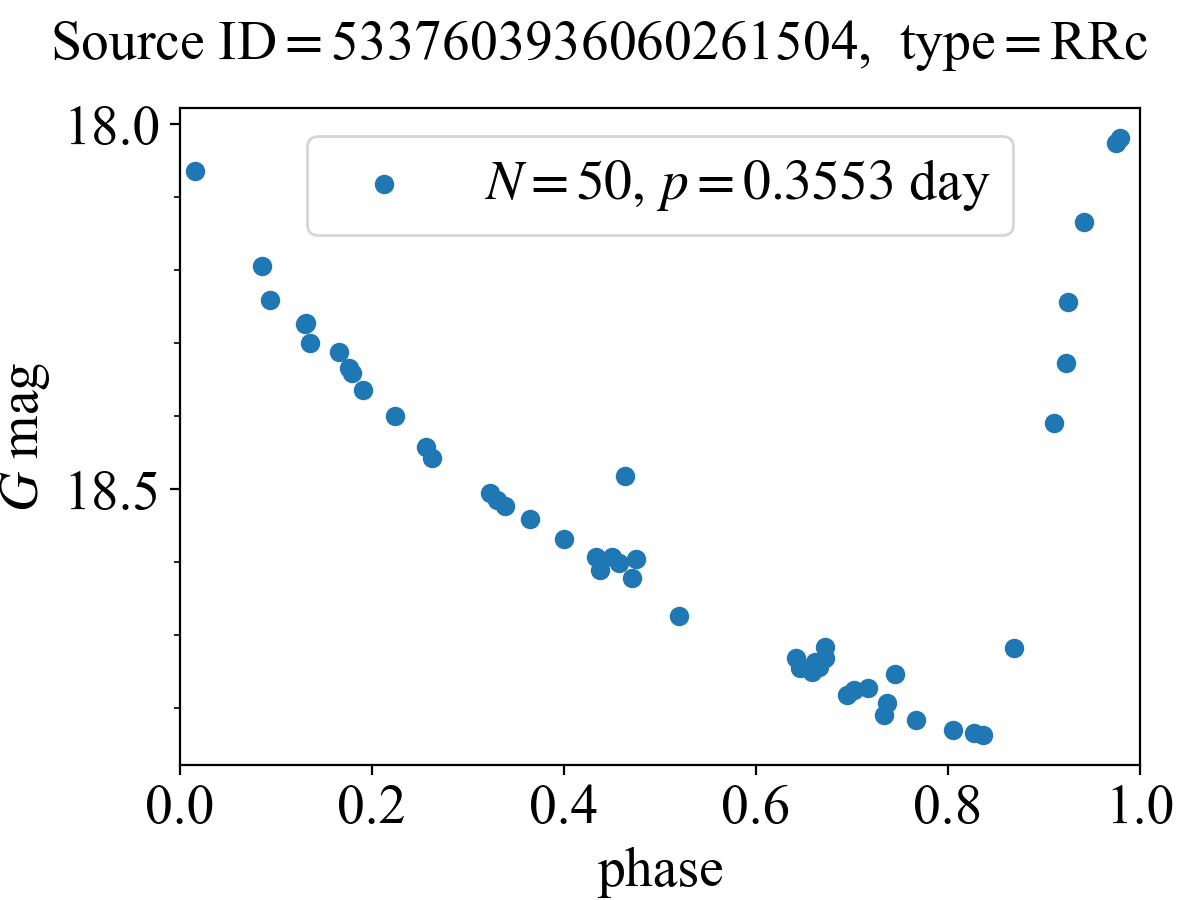}}
    \subfigure{\includegraphics[width=0.32\linewidth]{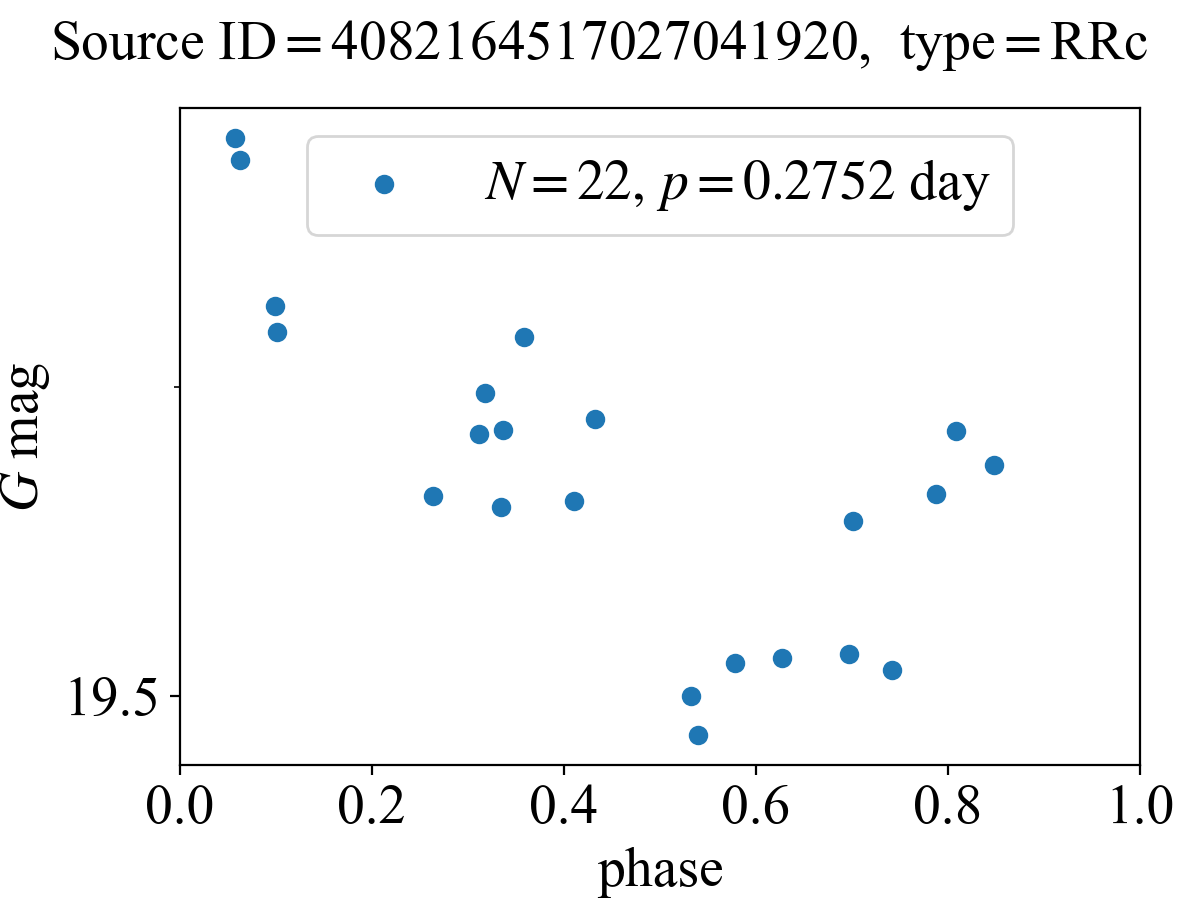}}
	\caption{\label{figbronze} Three representative light curves provided by Gaia DR3} from the bronze sample are shown, which should be carefully used. The light curve in the left panel appears too symmetric to be consistent with a typical RR Lyrae star, yet it is classified as an RRab. The middle panel shows a steep rise and an amplitude exceeding 0.8 mag, suggesting it should be classified as an RRab rather than an RRc. The light curve in the right panel contains few and scattered data points, making it difficult to reliably identify as an RR Lyrae star.
		\vspace{10pt}
\end{figure*}

First, we verify whether these RRLs can be confidently identified as RRLs and accurately classified into RRab/RRc subtypes. To achieve this, we utilize light curves collected by Gaia DR3 \citep{gaiacollaboration2022gaia}, the Zwicky Transient Facility Public Data Release \citep{Masci_2018}, and the Catalina Sky Survey \citep{2009ApJ...696..870D}. Our sample is cross-matched with these light curve catalogs, and we individually inspect the light curves of the selected RRLs. Focusing on light curve quality, we categorize the selected RRLs into three classes:

\begin{enumerate}
    \item[(1)] ``Gold sample": Contains at least one light curve observation that clearly matches the characteristic light curve pattern of an RRL subtype; 
    \item[(2)] ``Silver sample": Contains light curves with larger dispersion or fewer data points;
    \item[(3)] ``Bronze sample": Difficult to confidently identify as an RRL or potentially misclassified. The unreliable light curves of bronze sample can lead to unrealistic metallicities, requiring cautious use. Three typical example of bronze sample light curves are shown in Fig. \ref{figbronze}.
\end{enumerate}

\begin{figure}[h]
	\centering
	\includegraphics[width=0.6\linewidth]{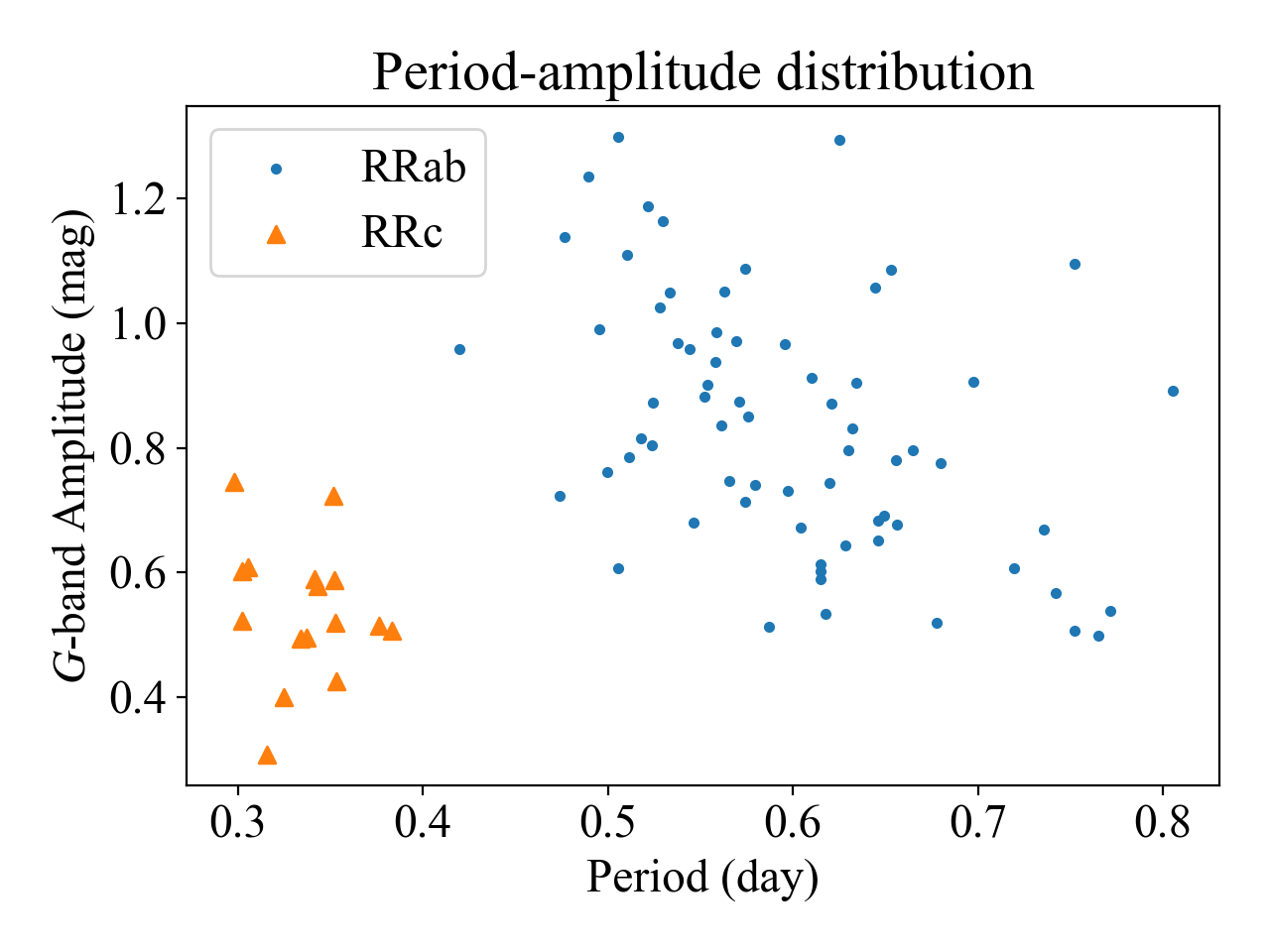}
	\caption{\label{figA} Period-amplitude distribution of the HV-RRL candidates for golden and silver samples. Blue circle dots are RRab types, orange triangles are RRc types.}
\end{figure}

It is particularly important to note that, RRab stars should have periods of $0.3\text{--}1.2$ days and amplitudes of $0.5\text{--}2$ mag, while RRc stars have periods of $0.2\text{--}0.5$ days and amplitudes of less than 0.8 mag \citep{1986Msngr..43...16M}. Additionally, periods provided by the catalogs should be consistent with the periods we derived from the light curves using Lomb-Scargle
Periodograms. RRLs that fail to meet the period-amplitude criteria should be classified as part of the bronze sample. Fig. \ref{figA} shows the period-amplitude distribution of the final confirmed gold and silver HV-RRL samples.


To summarize the examinations above, we identified 64 ``gold sample" HV-RRL candidates with reliable light curves, 23 ``silver sample" HV-RRL candidates with moderately reliable light curves and 78 ``bronze sample" HV-RRL candidates with doubtful light curves, which should be carefully used. A portion of the information for the selected 165 HV-RRL candidates is presented in Table \ref{tab_total}.

\begin{deluxetable*}{cccccccccccc}
\tabletypesize{\footnotesize}
\tablenum{2}
\tablecaption{\label{tab_total}Basic parameters for 165 HV-RRL candidates}
\tablehead{Gaia ID & RA & Dec & $\mu_{\alpha}\cos{\delta}$ & $\mu_{\delta}$ & $D$ & $R_{\rm GC}$ & $v_{{\rm GC},t}$ & Ref. & Quality \\
 & (deg) & (deg) & ($\rm mas\,yr^{-1}$) & ($\rm mas\,yr^{-1}$) & (kpc) & (kpc) & ($\rm km\,s^{-1}$) & & }
\startdata
378807525573579520 & $5.4851$ & $37.9437$ & $24.97\pm0.06$ & $-1.75\pm0.06$ & $14.1\pm1.5$ & $18.9\pm1.3$ & $1530.9\pm171.5$ & L23 & bronze\\
4918733303434183296 & $6.8229$ & $-57.2477$ & $1.29\pm0.27$ & $-1.78\pm0.25$ & $73.6\pm7.8$ & $71.4\pm7.6$ & $549.9\pm127.1$ & L23 & silver\\
4717299539114264320 & $22.8983$ & $-58.8057$ & $0.69\pm0.28$ & $-2.13\pm0.31$ & $69.7\pm6.4$ & $68.5\pm6.4$ & $574.8\pm122.5$ & L23 & gold\\
4910044167981920384 & $23.4723$ & $-56.0680$ & $1.18\pm0.22$ & $-1.36\pm0.27$ & $67.2\pm6.7$ & $66.2\pm6.7$ & $369.1\pm99.4$ & L23 & gold\\
4936275766641005312 & $30.3697$ & $-52.8000$ & $1.16\pm0.23$ & $-1.65\pm0.24$ & $68.7\pm6.5$ & $68.5\pm6.4$ & $462.0\pm102.3$ & L23 & gold\\
......\\
2004005081455404416 & $343.6118$ & $56.4564$ & $-1.94\pm0.32$ & $-1.79\pm0.32$ & $31.5\pm3.6$ & $34.9\pm3.5$ & $474.6\pm67.7$ & L23 & bronze\\
2644870582050682240 & $349.9803$ & $-0.1865$ & $11.48\pm0.11$ & $-9.32\pm0.11$ & $18.3\pm1.3$ & $19.3\pm1.1$ & $1092.5\pm86.4$ & W22 & bronze\\
2215178865230742912 & $351.7092$ & $70.3796$ & $0.92\pm0.34$ & $1.97\pm0.35$ & $48.1\pm4.9$ & $52.3\pm4.9$ & $410.8\pm93.3$ & L23 & gold\\
6522342764544281856 & $356.1458$ & $-52.6550$ & $0.89\pm0.21$ & $-2.11\pm0.27$ & $63.4\pm6.2$ & $60.8\pm6.1$ & $462.8\pm111.4$ & L23 & gold\\
6343475759127147008 & $356.6685$ & $-85.9455$ & $1.45\pm0.15$ & $-2.24\pm0.19$ & $43.9\pm4.3$ & $40.5\pm4.2$ & $383.9\pm66.1$ & L23 & gold\\
\enddata
\tablecomments{The complete table is available online.}
\end{deluxetable*}

\begin{deluxetable*}{ccccccc}
\tabletypesize{\footnotesize}
\tablenum{3}
\tablecaption{\label{tab_Punb} Unbound probabilities $P_{\rm unb}$ for different potential models}
\tablehead{Gaia ID & Bovy15 & Cautun20 & Monari18 & McMillan17 & Irrgang13I & Bovy15+LMC}
\startdata
378807525573579520 & 1.00 & 1.00 & 1.00 & 1.00 & 1.00 & 1.00\\
4918733303434183296 & 0.97 & 0.95 & 0.93 & 0.87 & 0.85 & 0.82\\
4717299539114264320 & 0.99 & 0.97 & 0.96 & 0.91 & 0.89 & 0.98\\
4910044167981920384 & 0.65 & 0.56 & 0.47 & 0.32 & 0.27 & 0.23\\
4936275766641005312 & 0.91 & 0.87 & 0.81 & 0.69 & 0.64 & 0.75\\
......\\
2004005081455404416 & 0.90 & 0.83 & 0.72 & 0.53 & 0.45 & 0.90\\
2644870582050682240 & 1.00 & 1.00 & 1.00 & 1.00 & 1.00 & 1.00\\
2215178865230742912 & 0.76 & 0.67 & 0.56 & 0.41 & 0.36 & 0.64\\
6522342764544281856 & 0.88 & 0.82 & 0.77 & 0.64 & 0.59 & 0.61\\
6343475759127147008 & 0.58 & 0.44 & 0.31 & 0.16 & 0.13 & 0.07\\
\enddata
\end{deluxetable*}

\section{Results} \label{sec:resul}
\subsection{Velocity distribution}\label{resul:v}  
\begin{figure*}
	\centering
	\includegraphics[width=1\textwidth]{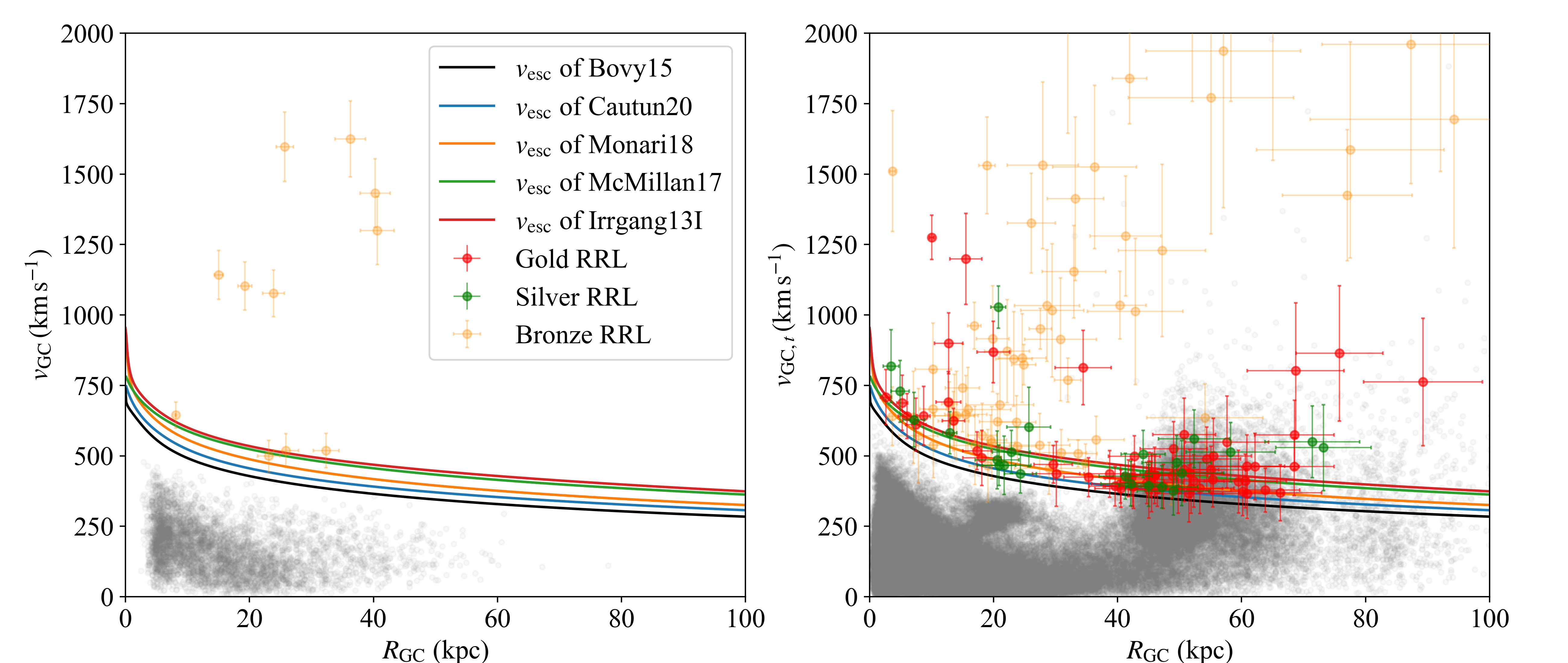}
	\caption{\label{figv}Velocity distribution of selected hypervelocity RRL candidates in the Galactic rest frame. Left panel shows RRLs with radial velocity; right panel shows RRLs without radial velocity. Red dots represent the HV-RRL candidates in the gold sample; green dots represent those in the silver sample; orange dots represent those in the bronze sample. Error bars are computed by Monte Carlo sampling. Grey dots depict the rest of the RRLs in the total sample. Black, blue, orange, green and red line denotes the escape velocity computed from \citet{2015ApJS..216...29B}, \citet{2020MNRAS.494.4291C}, \citet{2018A&A...616L...9M}, \citet{2017MNRAS.465...76M} and \citet{2013A&A...549A.137I}, respectively.}
		\vspace{10pt}
\end{figure*}

After selection and further examination, the final HV-RRL candidates are presented in Fig. \ref{figv}. Among the gold and silver samples, none of the HV-RRL candidates have measured radial velocities. Although many HV-RRL candidates are located just above the Milky Way’s escape velocity curve, we identify seven candidates in the gold and silver samples with tangential velocities exceeding 800 $\rm km\,s^{-1}$, including three with velocities surpassing 1000 $\rm km\,s^{-1}$.

We also find that HV-RRL candidates in the bronze sample predominantly exhibit exceptionally high velocities with larger error bars. This suggests that the questionable light curves of these candidates contribute to the uncertainty in their velocity measurements. It is important to note that in the following sections, we present results only for the gold and silver samples, which we consider more reliable.

In addition to \texttt{MWPotential2014} \citep{2015ApJS..216...29B}, we also employed other potential models of the Milky Way to calculate the escape velocity, including:
(1) \texttt{Cautun20} \citep{2020MNRAS.494.4291C}, a mass model fitted to Gaia DR2 rotation
curve of the Milky Way as measured by \citet{2019ApJ...871..120E};
(2) \texttt{Monari18} \citep{2018A&A...616L...9M}, a mass model fitted to the escape speed curve of the Milky Way based on the analysis of the velocity distribution of halo stars from the Gaia DR2;
(3) \texttt{McMillan17} \citep{2017MNRAS.465...76M}, which is an updated version of \citet{2011MNRAS.414.2446M}, utilizing observation data such as terminal velocity curve, maser and etc.
(4) \texttt{Irrgang13I} (model I from \citealt{2013A&A...549A.137I}), which is an updated version of \cite{1991RMxAA..22..255A}. These potentials all provided a higher escape velocity curve compared to \texttt{MWPotential2014}, as is displayed in Fig. \ref{figv}.

Finally, to consider the influence of the LMC, we modeled the LMC potential by Hernquist profile \citep{1990ApJ...356..359H} with a total mass of $1.38\times10^{11}\,M_\odot$ \citep{2019MNRAS.487.2685E} and a scale radius of 16 kpc to ensure the mass enclosed at
8.7 kpc matches the measured value of $1.7\times10^{10}\,M_\odot$ \citep{2014ApJ...781..121V}. As the fact that the LMC is orbiting the Milky Way with high speed, the stars associated with the LMC should have relative velocity compared to the LMC above the escape velocity of the LMC potential to be truely unbound. The unbound probabilities ($P_{\rm unb}$) under different potential models are summarized in Table \ref{tab_Punb}.

\subsection{Spatial distribution}

\begin{figure*}
	\centering
	\includegraphics[width=1\textwidth]{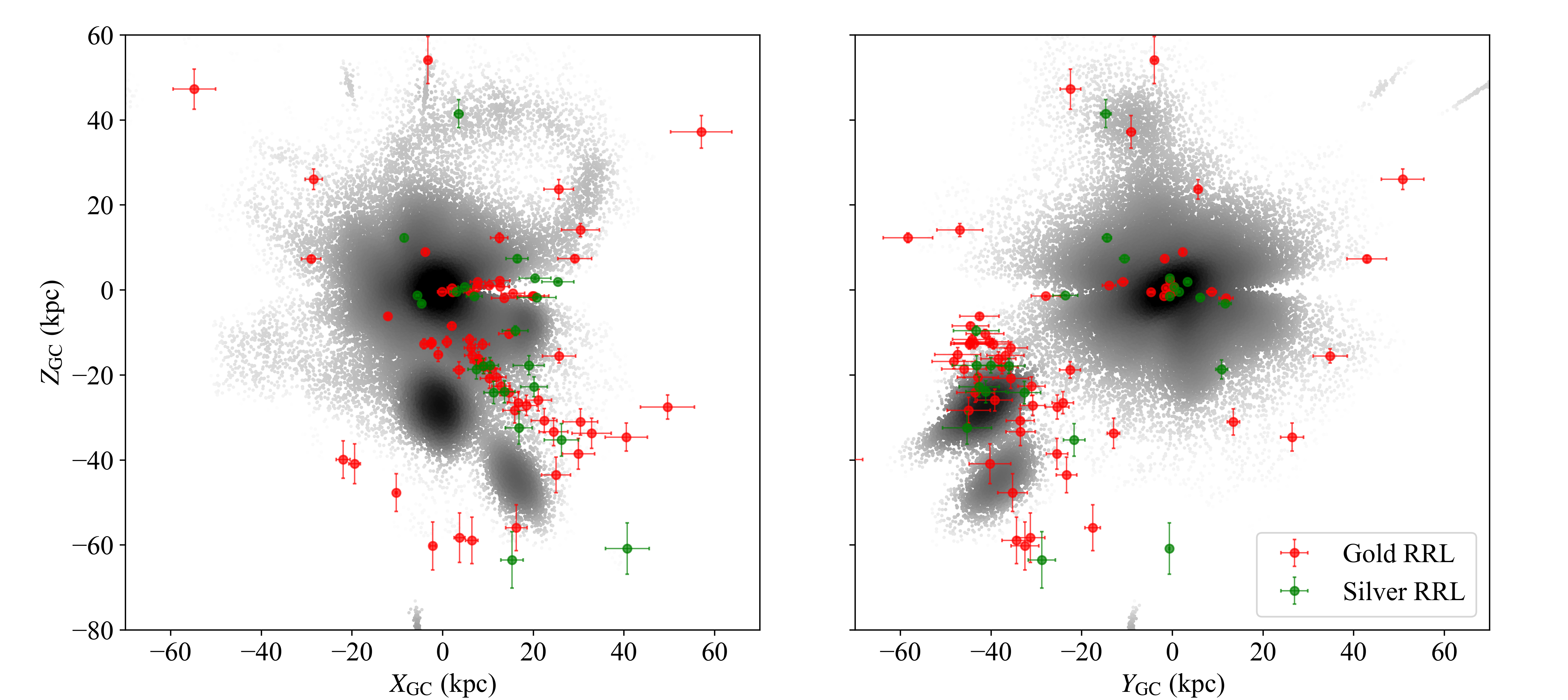}
	\caption{\label{figxyz} Spatial distribution of HV-RRL candidates in the gold and silver sample in the Galactic frame. Red dots represent the HV-RRL candidates in the gold sample; green dots represent in the silver sample. Error bars are computed by Monte Carlo sampling. Grey dots represent the rest of RRLs in the total sample, with depth of gray indicating logarithm of density.
	}
		\vspace{10pt}
\end{figure*}

\begin{figure*}
	\centering
	\subfigure{\includegraphics[width=0.75\linewidth]{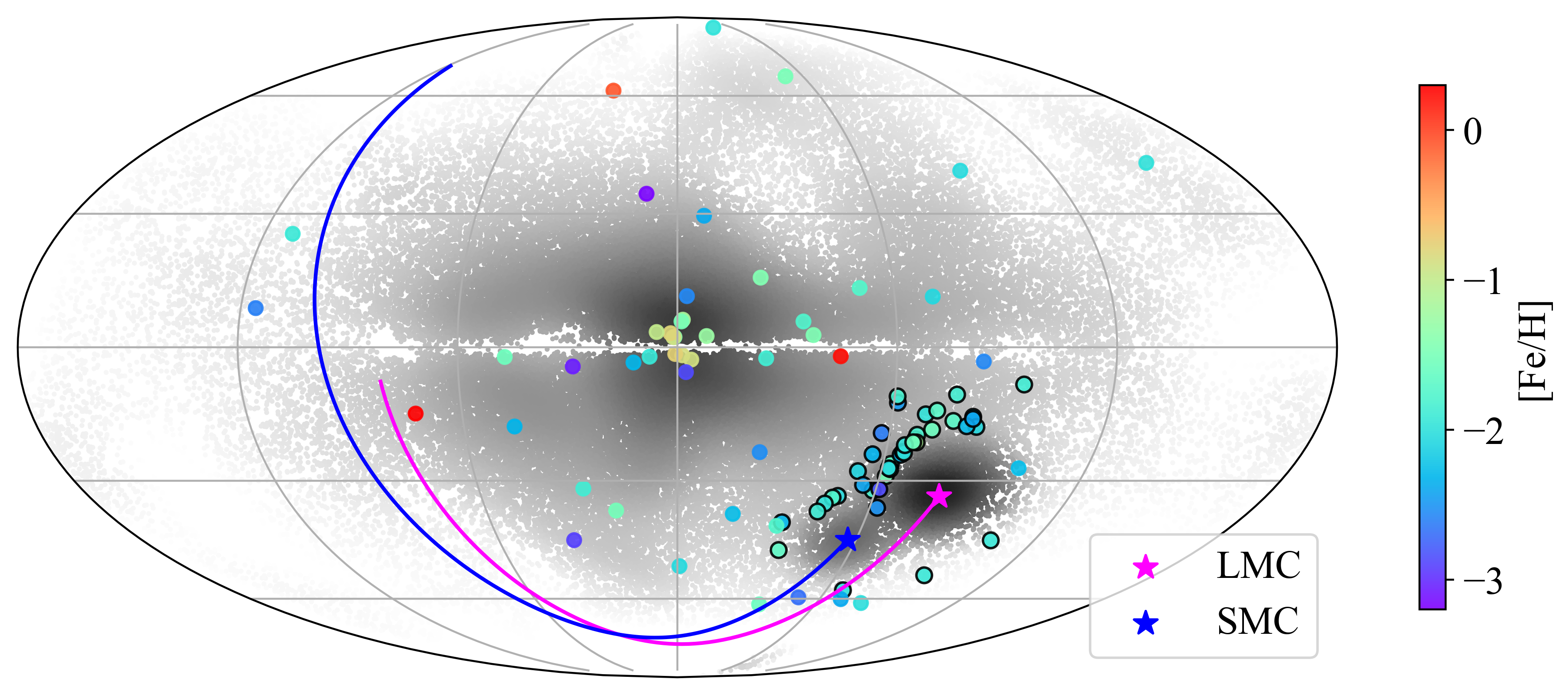}}
	\subfigure{\includegraphics[width=0.75\linewidth]{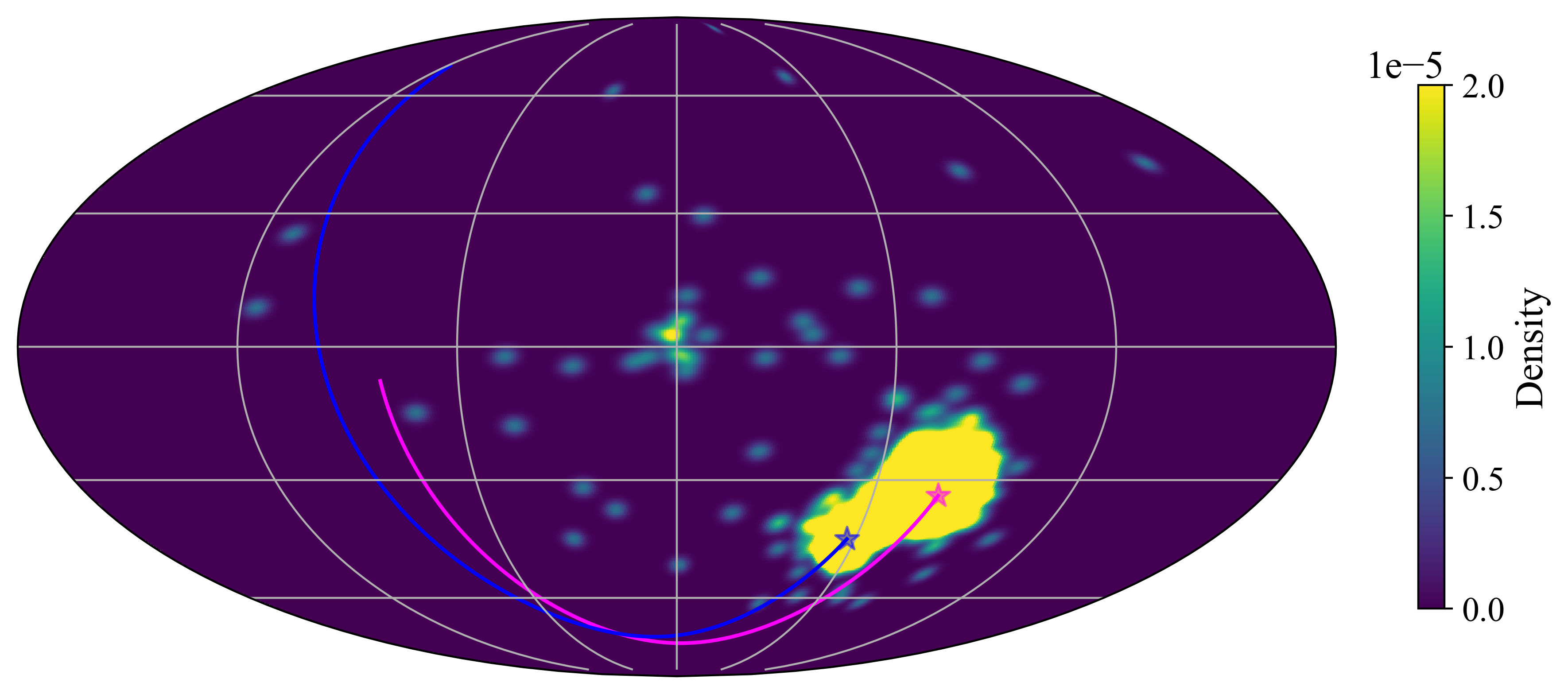}}\\
    \caption{\label{figlb} Spatial distribution of HV-RRL candidates in the Galactic coordinates using a Mollweide projection. In the upper panel, colors of the dots indicate the [Fe/H] of the HV-RRL candidates, with black circles indicating they are neighbors of the LMC and SMC. Grey dots represent the rest of RRLs in the total sample, with depth of gray indicating logarithm of density. Magenta and blue stars and lines denote the current position and previous orbit of the LMC and SMC respectively. The lower panel displays the density of RRLs with velocity exceeding the Milky Way's escape velocity, including the unbound member stars of the LMC and SMC.}
		\vspace{10pt}
\end{figure*}

The spatial distribution of HV-RRL candidates in the gold and silver sample is shown in Fig. \ref{figxyz}. Notably, some HV-RRL candidates appear to cluster around the LMC and SMC.

Additionally, we present a Mollweide projection of the spatial distribution in the Galactic coordinates $(l,b)$ in Fig. \ref{figlb}. In the upper panel, two distinct spatial distributions are evident: some HV-RRL candidates are concentrated toward the GC, while a larger number are localized around the LMC and SMC. 

These distributions are consistent with the findings of \cite{2021MNRAS.507.4997E}. By simulating the Hills mechanism originating from the SMBH at the GC and the center of the LMC, they demonstrated that HVSs ejected from the GC tend to remain in close proximity to the GC on the sky and are largely confined to the Galactic mid-plane. In contrast, HVSs ejected from the LMC will be a tight clump slightly leading the LMC’s orbit. Furthermore, the number of observable HVSs from the LMC is predicted to be significantly higher than those from the GC, which is also consistent with our results.

However, it remains uncertain whether our HV-RRL candidates were ejected via the Hills mechanism. Other processes, such as tidal stripping from satellite galaxies or runaway stars produced by supernovae in binary systems, may also contribute to their origin.
\cite{2017MNRAS.469.2151B} studied the runaway stars from the LMC through N-body simulations. They found that unbound runaway stars from the LMC are also centered around the LMC, with a particularly large number of stars exceeding the Milky Way's escape velocity within the LMC. In the lower panel of Fig. \ref{figlb}, approximately 7,000 unbound member stars of the LMC and SMC are included to illustrate a continuous density distribution. However, it should be noted that these unbound stars have not undergone light curve examination and may therefore be unreliable.
To further distinguish between the Hills mechanism and the runaway scenario, ejection velocity distributions should be considered, necessitating follow-up observations to measure radial velocities. 

\subsection{Metallicity distribution}

\begin{figure*}
	\centering
	\subfigure{\includegraphics[width=0.48\linewidth]{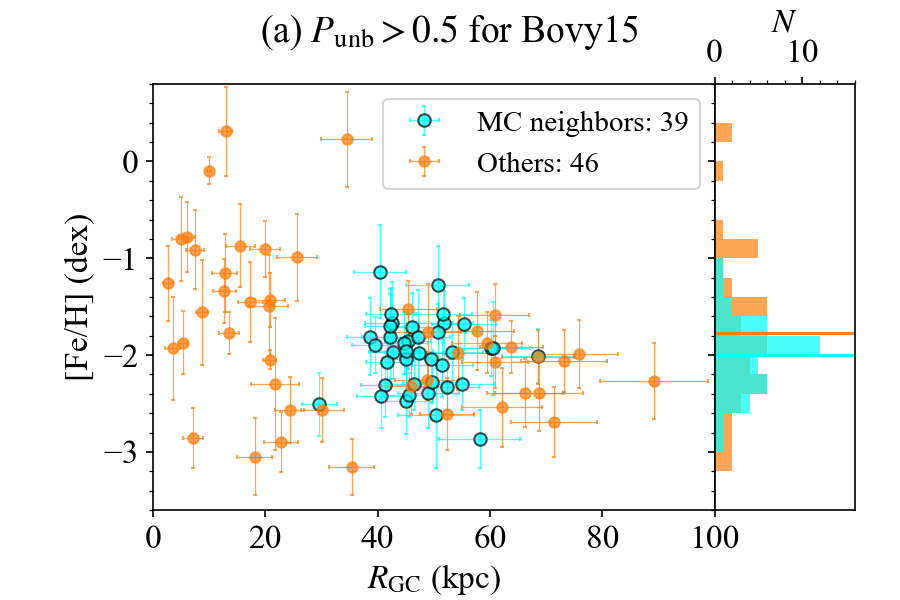}}
	\subfigure{\includegraphics[width=0.48\linewidth]{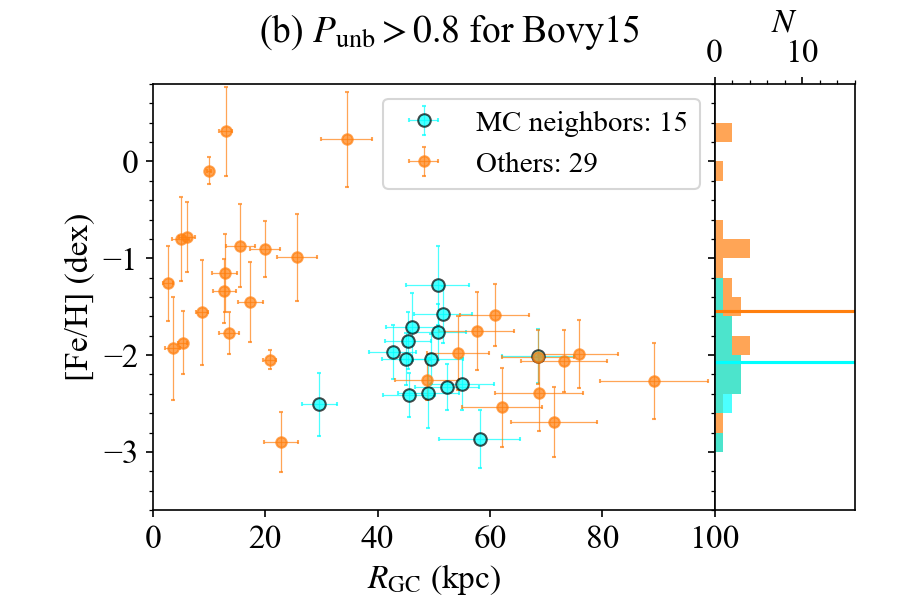}}\\
	\subfigure{\includegraphics[width=0.48\linewidth]{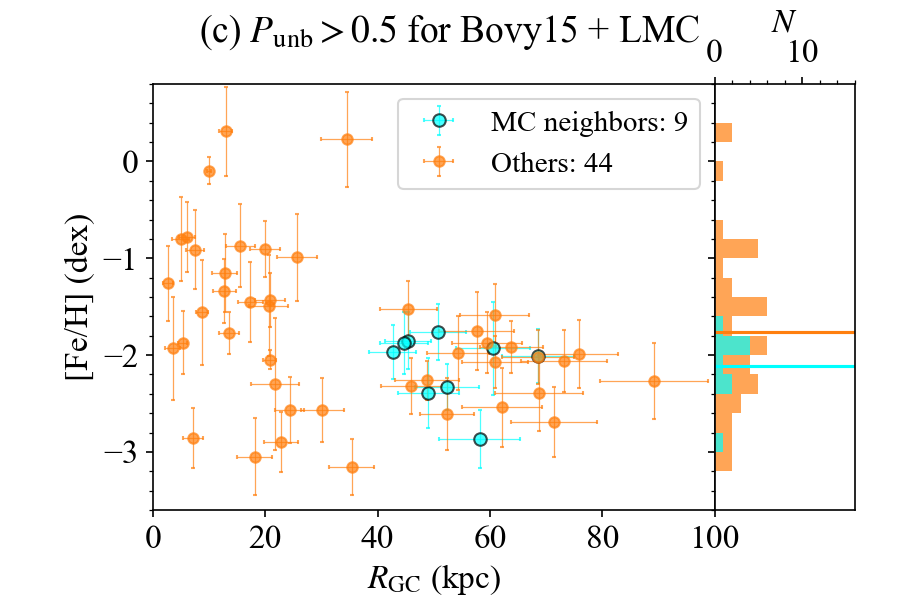}}
	\subfigure{\includegraphics[width=0.48\linewidth]{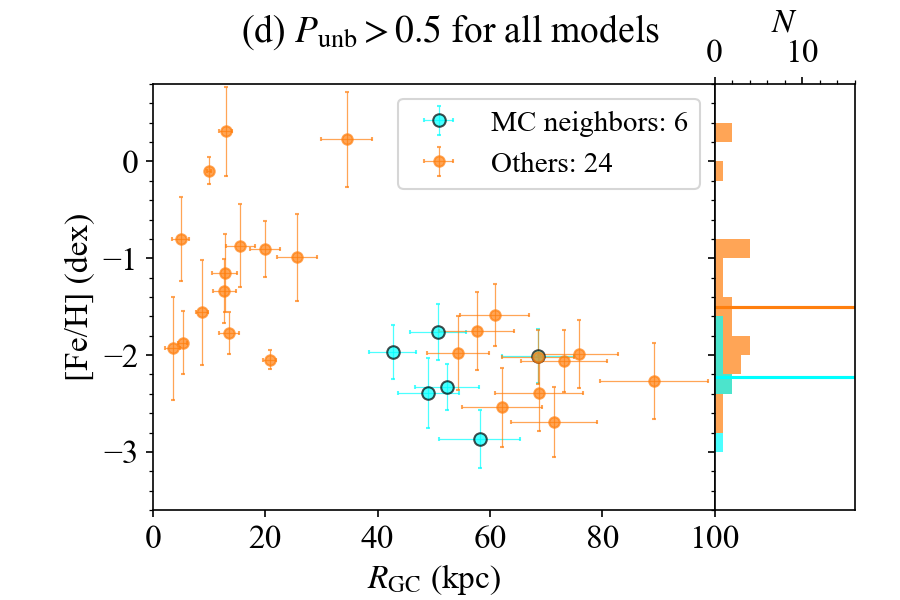}}
	\caption{\label{8check}Galactocentric distance versus metallicity distribution of the selected HV-RRL candidates. Cyan points with black outlines and error bars represent candidates located near the Magellanic Clouds, while orange points with error bars denote other candidates. The right subpanel of each main panel shows the corresponding [Fe/H] histogram, with horizontal lines marking the mean metallicity. Panels (a) through (d) illustrate the effects of varying selection criteria and gravitational potential models introduced in Section~\ref{resul:v}.}
		\vspace{10pt}
\end{figure*}

To further investigate the metallicity distribution, the [Fe/H] estimates of HV-RRL candidates are represented by the color of the dots in the upper panel of Fig.~\ref{figlb}, while the $R_{\rm GC}$--[Fe/H] relation is shown in Fig.~\ref{8check}. 
HV-RRL candidates located within 30\,kpc of the LMC and 20\,kpc of the SMC are identified as neighbors of the Magellanic Clouds (MC neighbors) and are marked with black-edged circles. 
The MC neighbors and other candidates in the Galactic halo typically exhibit metallicities around $\rm [Fe/H] \approx -2$, while HV-RRL candidates at smaller Galactocentric distances ($R_{\rm GC} < 20$\,kpc) tend to have higher metallicities, centered around $\rm [Fe/H] \approx -1.5$. This modest difference in metallicity distributions may point to different origins for the two populations. Additionally, MC neighbors exhibit similar metallicities compared with the mean values of RRLs associated with the LMC ($-1.67$) and SMC ($-1.95$) as inferred from our RRL sample.

We further evaluate the impact of different selection criteria and gravitational potential models in Fig.~\ref{8check}. Increasing the adopted escape velocity ($v_{\rm esc}$) reduces the overall number of HV-RRL candidates, but the two distinct spatial groupings remain clearly visible. Notably, Fig.~\ref{8check}(c), which includes the gravitational potential of the LMC, shows a significant reduction in the number of MC neighbors, while the rest of the HV-RRL population is largely unaffected. This implies that a substantial fraction of the initially identified MC neighbors are likely still gravitationally bound to the LMC.

\subsection{Orbit integration}

In addition to spatial and metallicity correlations, the most crucial approach to investigate the origin of HVSs is to trace their backward orbits. For HV-RRL candidates with complete 6D data, we calculated the historical minimum distance ($r_{\rm min}$) from the GC and Milky Way substructures, including 1,743 open clusters \citep{2021yCat..74785184D}, 145 globular clusters \citep{2010arXiv1012.3224H,2019yCat..74842832V,2021MNRAS.505.5978V,2021yCat..75055978V}, 39 satellite galaxies \citep{2018A&A...619A.103F}, as well as the LMC and SMC \citep{2020ApJ...893..121P}, with a total integration time of 50 Myr and a time step of 1 kyr (after a trial with total time of 1 Gyr and step of 100 kyr). We found one RRL (Gaia ID: 3745877687375105408) with a 19.0\% probability of having reached a backward-integrated minimum distance within 1 kpc of the GC.

However, most of our HV-RRL candidates lack radial velocity data, making orbit integration impossible without complete 3D velocity information. To address this, we assume the radial velocity follows a Gaussian distribution with a mean of 0 $\rm km\,s^{-1}$ and different dispersions of: 
\begin{itemize}
\item[(1)] 90~$\rm km\,s^{-1}$, corresponding approximately to the standard deviation of radial velocity of 688 thick disk stars in our RRL sample, defined by $z_{\rm max} < 4\,\rm kpc$, eccentricity $< 0.4$, and $L_z > 0$;
\item[(2)] 150~$\rm km\,s^{-1}$, representing the standard deviation of radial velocity of 3,823 remaining halo-like stars in our RRL sample;
\item[(3)] 300~$\rm km\,s^{-1}$, reflecting the expectation that HVSs originating from the GC may exhibit exceptionally high radial velocities.
\end{itemize} It is important to emphasize that this is a rough approximation, and the true radial velocities may differ from the assumed distributions. Nevertheless, this method offers a useful first-order insight into the possible origins of these stars. Using this approach, we identified one RRL (Gaia ID: 4055929177457043840) whose backward-integrated minimum distance falls within the half-light radius of the Sgr (2.6\,kpc; \citealt{2018A&A...619A.103F}). The associated probabilities under the three assumed radial velocity scenarios are 41.9\%, 40.9\%, and 37.4\%, respectively.

\section{discussion}\label{sec:discu}

\subsection{Radial velocity determination for J1300+0256}

None of the HV-RRL candidates in the gold and silver samples have measured radial velocities by spectroscopic surveys. However, to further validate our HV-RRL candidates and determine their origins, obtaining additional radial velocity measurements is essential.
Fortunately, we identified one candidate, J1300+0256 (Gaia ID: 3692326965681261184), with a radial velocity measurement in the recent Data Release 1 of the Dark Energy Spectroscopic Instrument \citep{2025arXiv250314745D}. J1300+0256 is an RRc-type star in the silver sample. Utilizing its metallicity of $-1.22\pm 0.18$ derived from the DESI spectrum with a mean SNR of 8.5, heliocentric distance is refined using the $M_G{\text{--}}{\rm [Fe/H]}$ relation for RRc stars provided by \citet{2023ApJ...944...88L}. 

The observed radial velocity of J1300+0256, derived from the H$\alpha$ absorption line in the DESI spectrum, is $163.6 \pm 12.2\,\rm km\,s^{-1}$.
However, due to the pulsation of RRLs, a correction is necessary to estimate the systematic radial velocity. 
To perform this correction, we employed the H$\alpha$ radial velocity template for RRc stars derived by \citet{2024RAA....24g5009H}.
Utilizing this template, the radial velocity curve of RRc stars can be reconstructed from Gaia $G$-band light curve amplitude and a single radial velocity measurement taken at a certain phase in the pulsation cycle.
Accordingly, a correction of $29.1 \pm 4.5\,\rm km\,s^{-1}$ is added to the observed radial velocity to derive the systematic radial velocity.
Incorporating the corrected radial velocity, the total velocity of J1300+0256 in the Galactic rest frame reaches $441.1\,\rm km\,s^{-1}$, exceeding the local escape velocity of $385.2\,\rm km\,s^{-1}$ at its position ($R=39.50\,\rm kpc$). With the full 3D velocity data, we integrated its past orbit but found no clear association with the GC. However, in 2000 Monte Carlo simulations, J1300+0256 can approach as close as 0.37 kpc to the globular cluster NGC 5824, which has a tidal radius of 0.25 kpc \citep{2020PASA...37...46B,2021MNRAS.505.5978V}.

\subsection{Comparison with previous study}

\cite{Prudil_2022} also conducted a search for HVSs using distances inferred from RRLs.
Their sample consisted of 6,187 RRLs compiled from Gaia DR2 \citep{2019A&A...622A..60C} and the Catalina Sky Survey \citep{2014yCat..22130009D}, all with 3D velocity measurements, from which they selected nine HV-RRL candidates.
Three of their candidates overlap with our study, corresponding to Gaia source IDs 3942539665718239360, 3692326965681261184, and 2644870582050682240. In our classification, the three stars fall into the gold, silver, and bronze categories, respectively.
Two of their candidates (Gaia IDs: 2471437298673558400 and 1018414089653402496) are not included in our initial sample. Compared to their study, we applied similar astrometric quality cuts to the Gaia data. Both studies used similar solar position and motion parameters and adopted the \texttt{MWPotential2014} model, with a selection criterion of $P_{\rm unb} > 0.5$. However, they did not apply the relative velocity uncertainty cut $\sigma_{v_{\rm GC}}/v_{\rm GC} < 0.3$, resulting in the exclusion of three of their candidates from our final selection (Gaia IDs: 3903031502808975616, 674694939356598528, and 3661050979472106880). Additionally, one of their candidates (Gaia ID: 3664033198603611520) is identified as an unbound member star of the Sgr in Section~\ref{2.3}. Therefore, it is not included in our final sample and will be discussed further in Section~\ref{dis:sub}.

\subsection{Member stars of the dwarf galaxies \label{dis:sub}} 

\begin{figure*}
	\centering
	\subfigure{\includegraphics[width=0.48\linewidth]{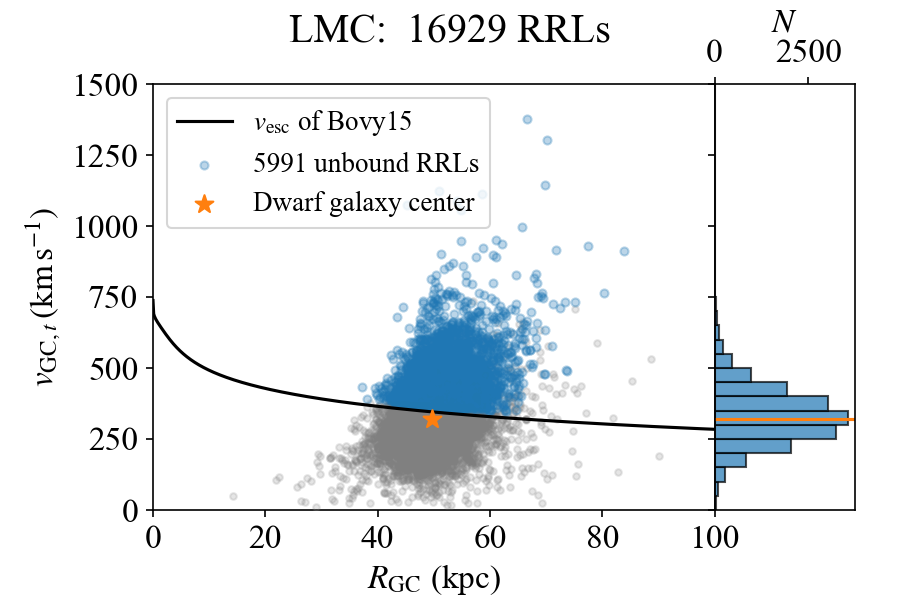}}
	\subfigure{\includegraphics[width=0.48\linewidth]{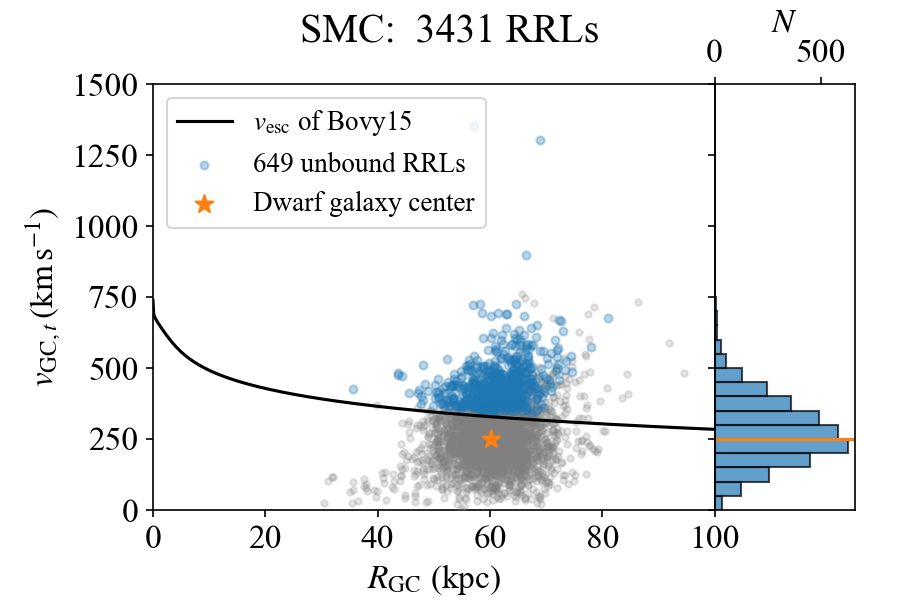}}\\
	\subfigure{\includegraphics[width=0.48\linewidth]{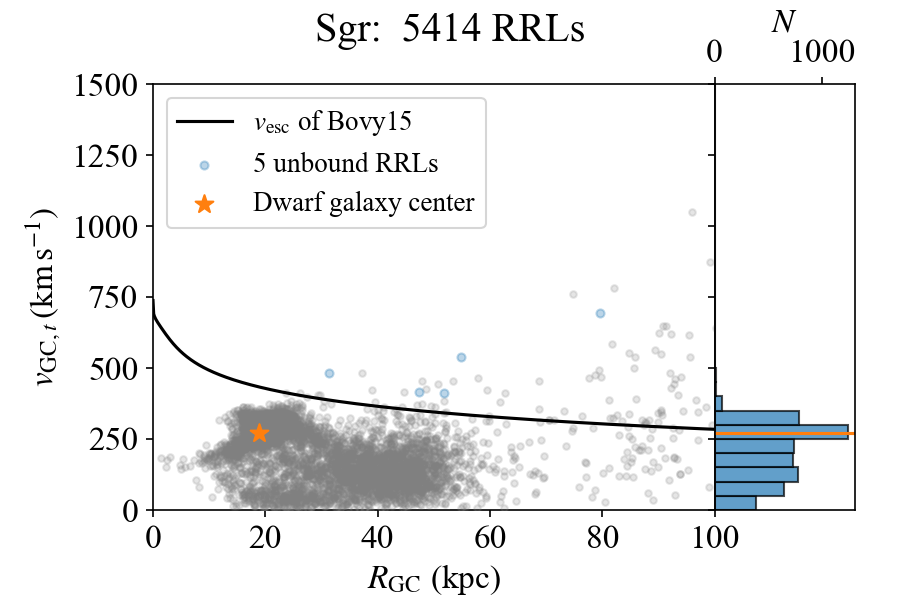}}
	\subfigure{\includegraphics[width=0.48\linewidth]{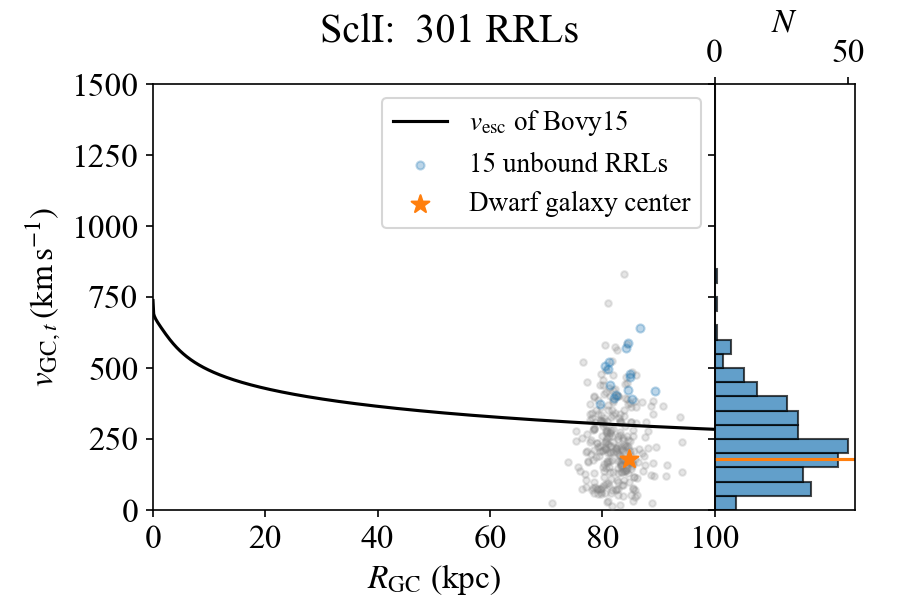}}\\
	\subfigure{\includegraphics[width=0.48\linewidth]{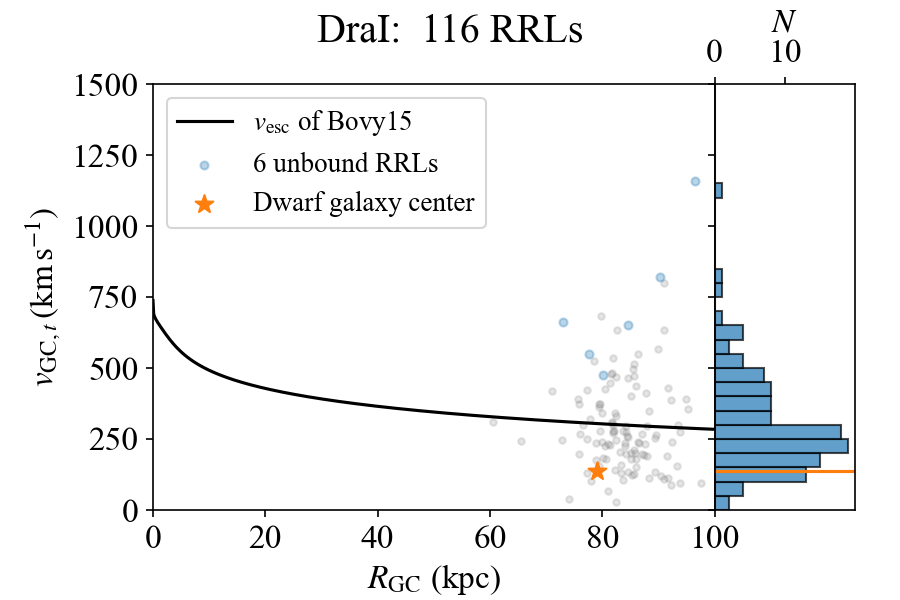}}
	\subfigure{\includegraphics[width=0.48\linewidth]{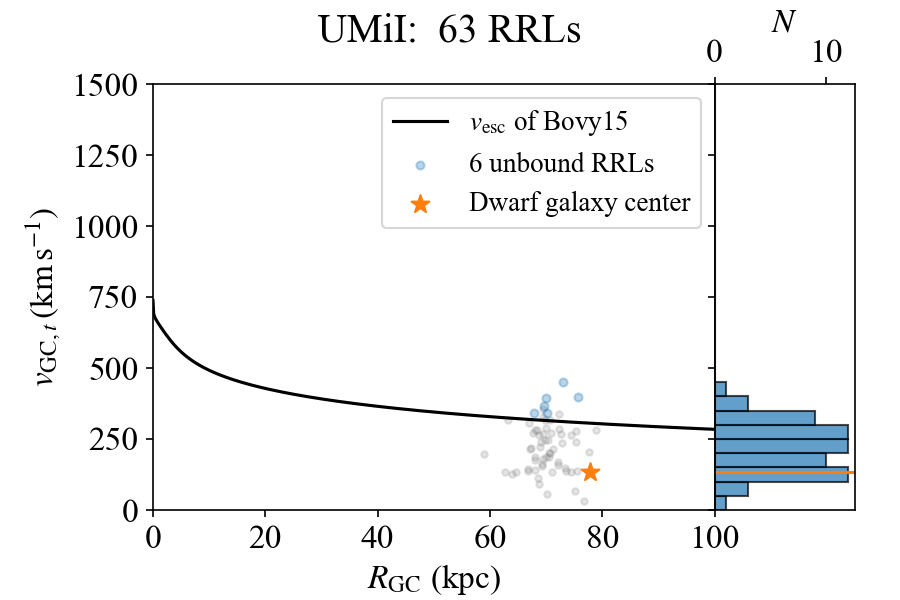}}
	\caption{\label{figsubs}The velocity distribution of the RRLs as member stars of the dwarf galaxies in the Galactic rest frame. Blue dots represent the selected unbound candidates of the Milky Way, while grey dots represent other RRLs. Orange stars mark the position and velocity of the dwarf galaxies' centers.}
		\vspace{10pt}
\end{figure*}

In this section, we further investigate the velocity distribution of RRLs as member stars of dwarf galaxies, selected from the total sample described in Section \ref{sec:metho}. In the Galactic rest frame, we find that some RRLs within the substructures have velocities significantly exceeding the escape velocity. Notably, for the LMC and SMC, a substantial number of RRLs lie above the escape velocity, as shown in Fig. \ref{figsubs}. This excess velocity may result from the high orbital velocities and tidal stripping near the pericenters of the dwarf galaxies' orbits \citep{2009ApJ...691L..63A}.

\begin{figure}
	\centering
	\includegraphics[width=0.7\linewidth]{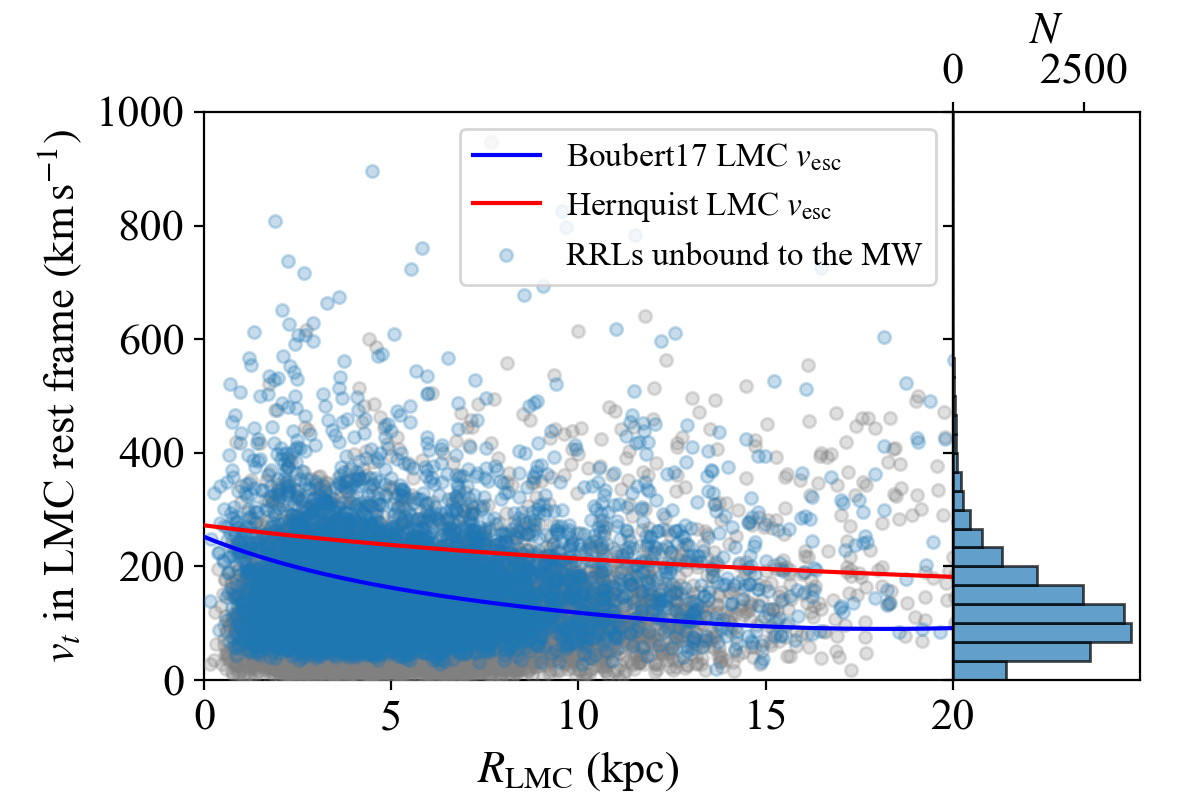}
	\caption{\label{figLMC}The velocity distribution of the RRLs as member stars of the dwarf galaxies in the LMC rest frame. Blue dots represent the selected unbound candidates of the Milky Way, while grey dots represent other RRLs. Blue line denotes the simulated LMC escape velocity in Fig. 4 from \cite{2017MNRAS.469.2151B}; Red line denotes the LMC escape velocity computed by Hernquist profile constructed in Section \ref{resul:v}.}
\end{figure}

To determine whether these stars are truly unbound from the dwarf galaxies and the Milky Way, we need to analyze their velocities in the rest frame of the dwarf galaxies. An example for RRLs in the rest frame of the LMC is shown in Fig. \ref{figLMC}, indicating that numerous stars can be unbound from both the LMC and the Milky Way. \cite{2017MNRAS.469.2151B} predicted there should be lots of stars escaping from the LMC through supernova in binary systems, with some also escaping the Milky Way's gravitational potential. This aligns with our observed results displayed in Fig. \ref{figLMC}.

Additionally, the Sculptor, Draco, and Ursa Minor dwarf galaxies (SclI, DraI, and UMiI) also exhibit a significant number of member stars with velocities exceeding the escape velocity, as shown in Fig. \ref{figsubs}. This phenomenon may be attributed to tidal stripping, as previous studies have shown that these dwarf galaxies can host extended stellar halos or tidal debris \citep{2022MNRAS.512.5601Q, 2024AJ....167...57T}. However, since these dwarf galaxies are currently far from their orbital pericenters, the contribution of tidal stripping to the production of HVSs may be less significant, as discussed by \cite{2009ApJ...691L..63A}. Furthermore, we highlight the possibility that these stars could be ejected via the Hills mechanism from central black holes within the dwarf galaxies. This is supported by the fact that some member stars exhibit velocities up to 800 $\rm km\,s^{-1}$ relative to the galaxy centers, which is comparable to the typical velocity produced by the Hills mechanism.

\section{Summary}\label{sec:summa}
Based on RRL catalogs from W22 and L23, combining astrometric measure by Gaia DR3, we identified 165 hypervelocity RR lyrae (HV-RRL) candidates. 

We've examined on their light curves, and categorized them into 64 ``gold sample", 23 ``silver sample" and 78 ``bronze sample" HV-RRL candidates to ensure we have reliable distance determination from RRL's light curve. The 87 HV-RRL candidates in the gold and silver sample have relatively more reliable light curves, while the bronze sample should be carefully used. Among the gold and silver sample, all the HV-RRL candidates have no radial velocity measurements, exceeding the escape velocity of the Milky Way by their tangential velocity. Notably, 7 RRLs have tangential velocity exceeding 800 $\rm km\,s^{-1}$, while 3 RRLs exceed 1000 $\rm km\,s^{-1}$.

Spatially, two distinct distributions are evident: one group is concentrated toward the GC, while a larger number are clustered around the LMC and SMC. This pattern is consistent with the simulations presented by \citet{2021MNRAS.507.4997E} and \citet{2017MNRAS.469.2151B}. Moreover, we observe a clear metallicity distinction between these two groups of candidates.

By integrating their backward orbits with a prior on radial velocity, we identified some HV-RRL candidates that could originate from the GC or the Sagittraus spheroid dwarf galaxy.
In addition, we identified a substantial number of RRLs as member stars of dwarf galaxies that exceed the Milky Way’s escape velocity, likely ejected from their hosts through various mechanisms.

In the future, we plan to conduct spectroscopic observations of HV-RRL candidates lacking radial velocity measurements. With these data, we will be able to more accurately determine their origins.

\section*{Acknowledgments}
This work is supported by the National Key R\&D Program of China No. 2024YFA1611903, the National Natural Science Foundation of China (NSFC grant Nos. 12090040, 12090044 and 12422303), Beijing Natural Science Foundation (no. 1242016) and the science research grants from the China Manned Space Project. This work was partially supported by Talents Program (24CE-YS-08) and the Popular Science Project (24CD012) of Beijing Academy of Science and Technology.

This work has made use of data from the European Space Agency (ESA) mission {\it Gaia} (\url{https://www.cosmos.esa.int/gaia}), processed by the {\it Gaia} Data Processing and Analysis Consortium (DPAC, \url{https:// www.cosmos.esa.int/web/gaia/dpac/ consortium}). Funding for the DPAC has been provided by national institutions, in particular the institutions participating in the {\it Gaia} Multilateral Agreement. 

This work has made use of data products from the Guo Shou Jing Telescope (the Large Sky Area Multi-Object Fibre Spectroscopic Telescope, LAMOST). LAMOST is a National Major Scientific Project built by the Chinese Academy of Sciences. Funding for the project has been provided by the National Development and Reform Commission. LAMOST is operated and managed by the National Astronomical Observatories, Chinese Academy of Sciences.

We thank Qingzheng Li and Xiangyu Li for their constructive assistance and appreciate Xinyi Li and Gaochao Liu for their guidance on data usage.

\bibliography{ref}{}
\bibliographystyle{aasjournal}

\clearpage
\end{document}